\documentclass[aps,prl,twocolumn,superscriptaddress,showpacs,amsmath,amssymb,longbibliography]{revtex4-1}
\usepackage[english]{babel}
\usepackage{amsmath}
\usepackage{bm}
\usepackage{graphicx,bbm} 
\usepackage{times}
\usepackage{epsfig} 
\usepackage{soul}
\usepackage[utf8]{inputenc}
\usepackage[colorlinks,linkcolor=blue,citecolor=blue,urlcolor=blue]{hyperref}
\usepackage{color}
\usepackage{latexsym}
\usepackage{ulem}
\definecolor{nred} {RGB}{224,0,0}
\definecolor{nblue} {RGB}{28,130,185}
\definecolor{dgreen} {RGB}{78,138,21}
\definecolor{norange}{RGB}{230,120,20}
\definecolor{mypurple}{RGB}{100,0,200}

\begin{document} 
\title{Spin diffusion in perturbed isotropic Heisenberg spin chain}
\author{S. Nandy}
\affiliation{Jo\v zef Stefan Institute, SI-1000 Ljubljana, Slovenia}
\author{Z. Lenar\v{c}i\v{c}}
\affiliation{Jo\v zef Stefan Institute, SI-1000 Ljubljana, Slovenia}
\author{E. Ilievski}
\affiliation{Faculty for Mathematics and Physics, University of Ljubljana, Jadranska ulica 19, 1000 Ljubljana, Slovenia}
\author{M. Mierzejewski}
\affiliation{Department of Theoretical Physics, Faculty of Fundamental Problems of Technology, Wroc\l aw University of Science and Technology, 50-370 Wroc\l aw, Poland}
\author{J. Herbrych}
\affiliation{Department of Theoretical Physics, Faculty of Fundamental Problems of Technology, Wroc\l aw University of Science and Technology, 50-370 Wroc\l aw, Poland}
\author{P. Prelov\v{s}ek}
\affiliation{Jo\v zef Stefan Institute, SI-1000 Ljubljana, Slovenia}

\date{\today}
\begin{abstract}
The isotropic Heisenberg chain represents a particular case of an integrable many-body system exhibiting superdiffusive spin transport at finite temperatures. Here, we show that this model has distinct properties also at finite magnetization $m\ne0$, even upon introducing the SU(2) invariant perturbations. Specifically, we observe nonmonotonic dependence of the diffusion constant ${\cal D}_0(\Delta)$ on the spin anisotropy $\Delta$, with a pronounced maximum at $\Delta =1$. The latter dependence remains true also in the zero magnetization sector, with superdiffusion at $\Delta=1$ that is remarkably stable against isotropic perturbation (at least in finite-size systems), consistent with recent experiments with cold atoms.
\end{abstract}

\maketitle

\noindent {\it Introduction.} 
The integrable quantum many-body lattice models and their anomalous finite-temperature $T>0$ transport properties have been the subject of theoretical investigations for many decades. The development of efficient analytical and numerical techniques \cite{bertini21} has recently significantly advanced our understanding. In this regard, the one-dimensional isotropic Heisenberg model played a prominent role. In spite of its exact solvability \cite{bethe31}, understanding the plethora of anomalous transport properties continues to posit a challenge, especially concerning the observed anomalous superdiffusive spin transport at $T>0$ emerging at the junction of the gapless regime at $\Delta <1$, implying finite spin stiffness $D(T>0)>0$ \cite{zotos99,prosen11,prosen13} featuring ballistic spin transport \cite{castella95,zotos96,zotos97}, and gapped regime for $\Delta>1$ with vanishing $D(T >0)=0$ and finite (dissipationless) diffusion constant ${\cal D}_0 < \infty$ \cite{znidaric11}. While postulated earlier \cite{fabricius98}, the superdiffusion of isotropic $\Delta=1$ point with particular universal Kardar-Parisi-Zhang (KPZ) dynamical scaling exponent $z=3/2$ has been recently established both numerically \cite{znidaric11,ljubotina17,ljubotina19,dupont20} and analytically within the generalised hydrodynamics (GHD) \cite{ilievski18,gopalakrishnan19,bulchandani20,bulchandani21,ilievski21}. It is worth noting that isotropic Heisenberg chains can be approximately realized in spin-chain materials \cite{motoyama96,scheie21} (possessing very large thermal conductivity owing to nearly conserved energy current \cite{hess07}), as well as in cold-atom optical lattices \cite{hild14,jepsen20,wei22}.

Although there are still open questions in integrable lattice models, the understanding of the effects of (even weak) integrability breaking perturbations (IBP) remains particularly challenging \cite{bastianello21,mallayya19,lange18}. In connection with the dc spin conductivity $\sigma_0$ and related spin diffusion ${\cal D}_0$, the role of the uniform (preserving translational symmetry) perturbative term $g H'$ has been addressed numerically within the easy-plane regime $\Delta < 1$ \cite{jung07,jung107,znidaric20,mierzejewski22}. It was proposed, via perturbation-theory arguments, that at high-$T$ and weak $g \ll 1$ the dc conductivity scales as $\propto 1/g^2$, but in general exhibiting multiple relaxation times \cite{mierzejewski22} related to the different conserved quantities involved in the current relaxation. On the other hand, in the easy-axis $\Delta >1$ regime, the role of IBP is unusual \cite{denardis21a,prelovsek22} due to finite, anomalous/dissipationless diffusion even in the integrable model. The spin transport in the perturbed isotropic Heisenberg model seems to be even richer. In particular, due to the SU$(2)$ spin symmetry, the isotropic perturbations that preserve such symmetry are expected to have different (even singular) effect on spin transport \cite{denardis21,wei22,roy22}, in contrast to anisotropic ones \cite{znidaric20}. 

In this Letter, we show that the distinctive transport properties of the isotropic Heisenberg model are, at high-$T$, not exclusive to the integrable (i.e., unperturbed) model or to vanishing magnetization density $m=0$, but are instead seen also at finite $m>0$ and finite isotropic perturbation strengths. We present numerical evidence that:  (i) In the $m=0$ sector for $\Delta=1$, superdiffusive transport is extremely robust to the isotropic perturbations of even moderate strength $g$,  exhibiting superdiffusive scaling of the diffusion constant $\mathcal{D}_0 \sim L^\zeta$ with $0<\zeta <1/2$ for system sizes up to $L=100$. (ii) For isotropic perturbations at $\Delta=1$, the diffusion constant ${\cal D}_0(\Delta)$ features a peak at $\Delta=1$ in all magnetization sectors. In contrast, the anisotropic IBP lead to a monotonic dependence ${\cal D}_0(\Delta)$. Away from $\Delta \simeq 1$ and $m=0$, in both cases, the results appear close to the standard perturbation theory ${\cal D}_0 \propto 1/g^2$ scaling, while for $\Delta=1$ the $g$ dependence of our results is less conclusive. (iii) Already in the unperturbed isotropic spin chain, the high-$T$ spin stiffness, measured in the units of static spin susceptibility, reveals a roughly linear dependence $D^*(m) \simeq 2|m|$ across a broad range of densities $m\gtrsim 0.3$ that have gone unnoticed so far. This dependence eventually crosses over to the non-analytic behavior $D^*(m) \simeq m^2 \log(1/|m|)$ at small $m$, which is hard to observe numerically. Finite magnetization results are obtained with the microcanonical Lanczos method (MCLM) on systems up to $L=36$ sites. However, the most challenging regime appears in the vicinity of critical point $m = 0, \Delta=1$. Here, we employ the time-evolving block decimation (TEBD) technique for boundary driven open systems with up to $L=100$ sites to establish the nonequilibrium steady state (NESS).

\noindent{\it Model.}
We study the $S=1/2$ XXZ Heisenberg spin chain with general anisotropy $\Delta$ adding the IBP of the strength $g$
\begin{equation}
H= J \sum_{i} \bigl[ \frac{1}{2} ( S^+_{i+1} S^-_i + \mathrm{H.c.} ) + 
\Delta S^z_{i+1} S^z_i \bigr] + g J H^{\prime}\,.
\label{xxz}
\end{equation}
We deal with the IBP, which (at least partly) conserve the translational symmetry of the model, and concentrate on the case of the staggered exchange. The anisotropic IBP contains only the spin-flip term \mbox{$H^{\prime}_{\mathrm{an}} = (1/2) \sum_i (-1)^i (S^+_{i+1} S^-_i + \mathrm{H.c.})$}, while as the alternative case we consider \mbox{$H^\prime_\mathrm{is}= H^\prime_{\mathrm{an}}+ \Delta \sum_i (-1)^i S^z_{i+1} S^z_i$} which is fully SU(2) isotropic for $\Delta =1$. We consider Hamiltonians that conserve total magnetization, namely $s^z_\mathrm{tot}= m L$, where $|m| \leq 1/2$ is the magnetization density of the system, and the spin current operator is given by \mbox{$j_s= (J/2) \sum_i [1 + g (-1)^i] (i S^+_{i+1} S^-_i + \mathrm{H.c.}) $.} We use $J=1$ as the unit of energy and analyse finite systems with $L$ sites, either closed with periodic boundary conditions (PBC) or open with boundary driving.

\noindent{\it Perturbed Heisenberg chain - finite magnetization.}
We first consider the high-$T$ dynamical conductivity $\tilde \sigma(\omega)= T \sigma(\omega)$ of a perturbed system with PBC and fixed magnetization $s^z_\mathrm{tot}= Lm$, focusing here on the intermediate $m=1/4$. We evaluate $\tilde \sigma(\omega)$ using the MCLM \cite{long03,prelovsek13,prelovsek21} by employing a large number of Lanczos steps, i.e., here typically $M_L = 20000$ for systems up to $L = 36$ with $N_{st} \lesssim 10^7$ basis states. This allows for frequency resolution $\delta \omega \simeq 10^{-3}$, important to resolve also the large dc $\tilde \sigma_0 = \tilde \sigma(\omega \to 0)$ emerging due to long relaxation times $\tau \gg 1$. Consequently, the dc diffusion constant can be extracted assuming the generalized Einstein relation, ${\cal D}_0= \tilde \sigma_0/(T \chi_0)$, \mbox{$\chi_0 =(1/\pi) \int \sigma(\omega) \mathrm{d}\omega = (1/4-m^2)/T$}, which is valid in perturbed/normal systems \cite{prelovsek22}. Note that the diffusion ${\cal D}_0$ is a well defined concept even at $T \to \infty$, unlike conductivity $\sigma_0=\sigma(\omega \to 0) \propto 1/T$.

\begin{figure}[tb]
\includegraphics[width=1.0\columnwidth]{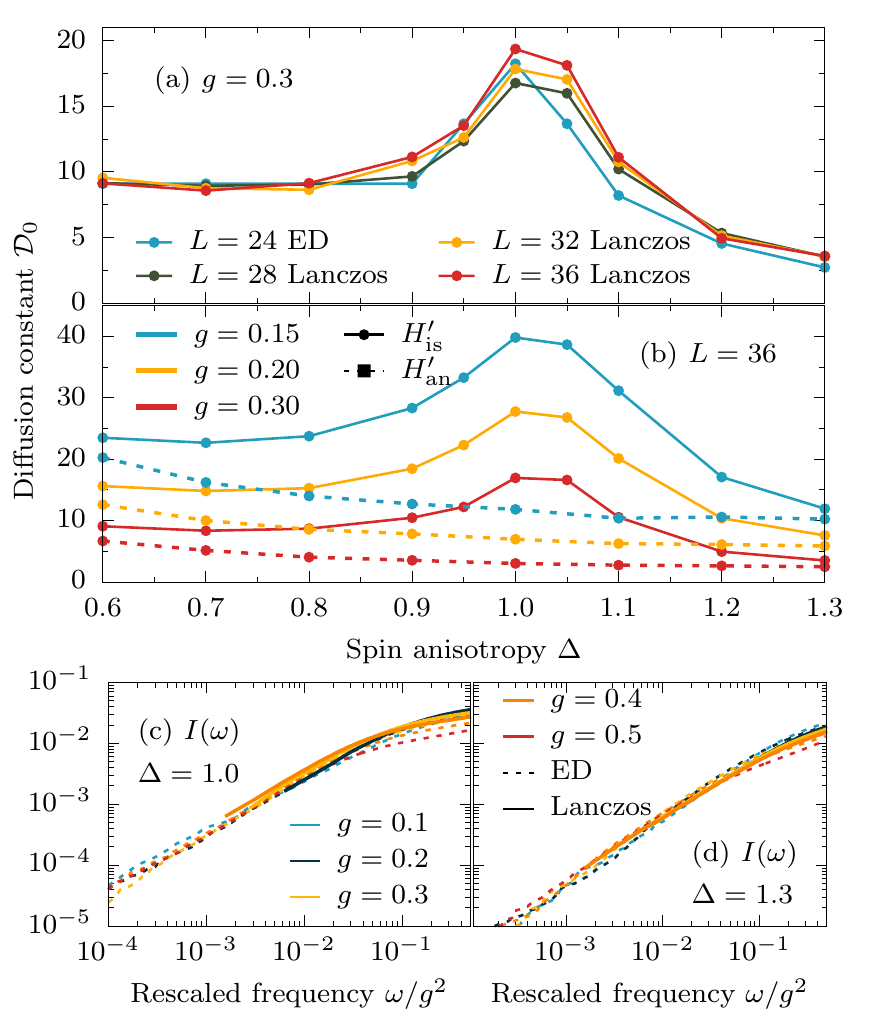}
\caption{Diffusion constant ${\cal D}_0$ vs. anisotropy $\Delta$ at fixed magnetization density $m=1/4$ for isotropic-perturbation strength $g=0.3$, as obtained for different system sizes via ED ($L=24$) and MCLM ($L= 28 -36$). (b) ${\cal D}_0$ vs. $\Delta$ for isotropic (full lines) and anisotropic (dashed lines) $g =0.15 - 0.3$, obtained via MCLM for $L=36$. (c,d) Integrated optical conductivity $I(\omega)$ for (c) $\Delta=1$ and (d) $\Delta=1.3$ as the function of the rescaled frequencies by the square of the perturbation strength $\omega/g^2$. Dashed (solid) line depict $L=20$ ($L=36$) ED (MCLM) data.
}
\label{fig1}
\end{figure}

Results for the diffusion constant ${\cal D}_0$ at magnetization density $m=1/4$ as the function of the anisotropy $\Delta$ are presented in Fig.~\ref{fig1}(a,b). Most strikingly, as shown in Fig.~\ref{fig1}(a), ${\cal D}_0$ reveals a pronounced peak at isotropic point $\Delta=1$ with high value ${\cal D}_0 \gg 1$ even at substantial $g=0.3$. The results are reliable despite the very large ${\cal D}_0$, also implying a narrow peak in dynamical $\tilde \sigma(\omega)$, as evidenced by analyzing systems of different numerical complexity $L =24 -36$. Dependence on the perturbation strength $g$ is quite consistent with expected standard ${\cal D}_0 \propto 1/g^2$, see Fig.~\ref{fig1}(b). In addition, in Fig.~\ref{fig1}(c,d), we present the frequency $\omega$ dependence of the integrated conductivity $I(\omega)=(1/\pi) \int_0^{\omega} \sigma(\omega^\prime) \mathrm{d}\omega^\prime$, from which the diffusion constant can be extracted as $I(\omega\to0)\propto {\cal D}_0\omega$. We observe the collapse of different $g$-curves, plotted as a function of frequencies rescaled by the square of the perturbation strength $\omega/g^2$. On the other hand, in Fig.~\ref{fig1}(b), we also display the effect of the anisotropic IBP $H^{\prime}_\mathrm{an}$, which behaves regularly in several respects: (i) the variation with $\Delta$ is smooth and monotonously decreasing without any specific feature at $\Delta=1$; (ii) ${\cal D}_0$ value is much smaller, at least at $\Delta \simeq 1$; (iii) the scaling with ${\cal D}_0 \propto 1/g^2$ can be accurately followed for all the considered values of $g$.

\begin{figure}[tb]
\includegraphics[width=1.0\columnwidth]{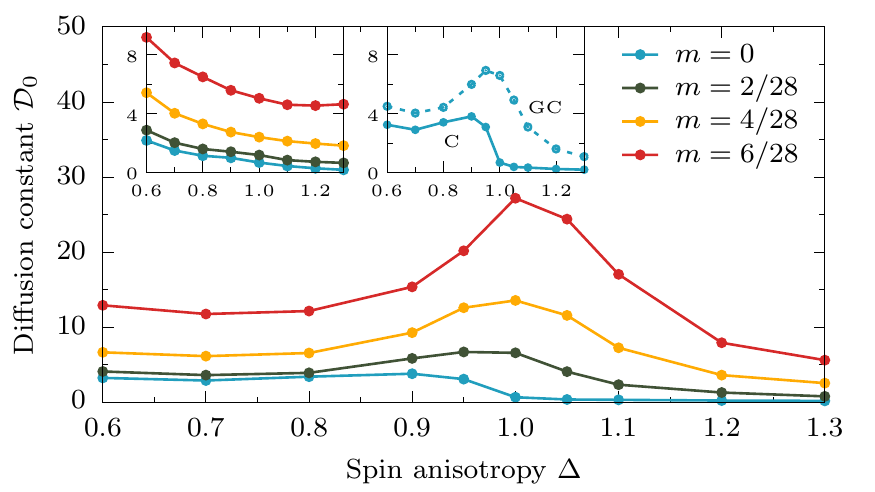}
\caption{Diffusion constant ${\cal D}_0$ vs. anisotropy $\Delta$, as calculated via MCLM in ensembles with magnetization densities $m= [0,6]/28$ and $g=0.2$ for isotropic (main panel) and anisotropic (left inset) perturbation, respectively. Right inset: comparison of canonical $m=0$ and grandcanonical results.} 
\label{fig2}
\end{figure}

In Fig.~\ref{fig2} we present also the variation of ${\cal D}_0(\Delta)$ as obtained via MCLM at fixed size $L=28$ in canonical ensembles with fixed magnetization $s^z_\mathrm{tot}= m L$, $m =[0,6]/28$, shown here for moderate $g=0.2$. The anisotropic IBP case (in the inset) again reveals a steady decrease of ${\cal D}_0$ with $\Delta$ for all $m$. By contrast, the isotropic IBP in the vicinity of $\Delta \simeq 1$ reveals a strikingly different variation of ${\cal D}_0$ with $m$, with $\mathcal{D}_0(\Delta)$ developing a peak around $\Delta=1$ with increasing $m>0$.

\noindent{\it Perturbed Heisenberg chain - zero magnetization.}
Our closed-system MCLM analysis indicates that the canonical $m=0$ results at accessible system sizes $L\le 28$ do not match with the corresponding grandcanonical average (see inset of Fig.~\ref{fig2} and \cite{supmat}). This remains true even for substantial $g \simeq 0.3$ and is most pronounced for isotropic IBP. As a consequence, the analysis of the spin transport in the vicinity of $m=0$ and $\Delta=1$ requires special attention, and previous studies have already reported that this regime is particularly sensitive to finite size/time effects \cite{denardis21}. To achieve larger $L$, we study open Heisenberg chains, where the spin current is driven via boundary Lindblad operators $L_1=\sqrt{1+\mu}S_{1}^{-}$, $L_2 = \sqrt{1-\mu}S_{1}^{+}$, $L_3= \sqrt{1-\mu}S_{L}^{-}$, $L_4 = \sqrt{1+\mu}S_{L}^{+}$ with a small spin bias $\mu$. To reach the non-equilibrium steady state (NESS) current $j_{ss}$ we use the time-evolving block-decimation technique for vectorized density matrices \cite{verstraete04,zwolak04}. In such a setup for diffusive systems, one finds a linear magnetization profile emerging at late times, and the spin diffusion constant can be extracted via ${\cal D}_0 = -j_{ss}/\nabla s^z$, where $\nabla s^z$ is the magnetization gradient \cite{znidaric11,supmat}. Ballistic transport corresponds to an essentially flat steady-state magnetization profile in the bulk, $s^z_i \equiv \textrm{tr}(S^z_i \rho_{ss})=0$,  with $s^z_2-s^z_1=s^z_N-s^z_{N-1}=\frac{\mu}{2}$ jump at the edges \cite{znidaric11,supmat}. For superdiffusive transport, spin profile resembles $s_i^z \simeq \frac{\mu}{\pi} \arcsin(-1+2\frac{i-1}{L-1})$ \cite{znidaric11,supmat}. In both of these cases, realized by $\Delta<1$ and $\Delta=1$, respectively, adding IBP should again give way to diffusive transport with a linear spin profile. While the bulk spin profile is indeed linear, in finite systems with IBP and $\Delta<1$, a ballistic component transversing the system without scattering yields a small jump at the edges, diminishing with increasing $L$ or $g$. For anisotropic IBP and $\Delta=1$, such jumps at the edges are almost negligible. For isotropic IBP and $\Delta=1$, on the other hand, an approximately linear bulk spin profile retains some curvature, while the jump at the boundaries is reminiscent of (or even more pronounced than in) the unperturbed $\arcsin(x)$ profile for finite $L$ \cite{supmat}. This already indicates that for isotropic IBP and $\Delta=1$, at considered system sizes $L \le 100$, there is still no restoration of normal transport. We calculate the diffusion constant from the magnetization gradient extracted from the $f=0.2$ fraction of the central bulk spin profile in all cases.

Anisotropic IBP results far away from $\Delta =1$, e.g. at $\Delta =0.5$, corroborate the expected scaling ${\cal D}_0 \propto 1/g^2$ \cite{supmat}. For $\Delta=1.0$ and anisotropic IBP, our results show normal $1/L$ finite size corrections, however, we cannot access small enough IBP strengths $g$, and consequently large enough $L$, to reveal the anticipated ${\cal D}_0 \propto 1/g^{2/3}$ scaling \cite{znidaric20}; for $g$ and $L$ parameters considered, we see no asymptotic scaling with $g$ yet \cite{supmat}.

\begin{figure}[tb]
\includegraphics[width=1.0\columnwidth]{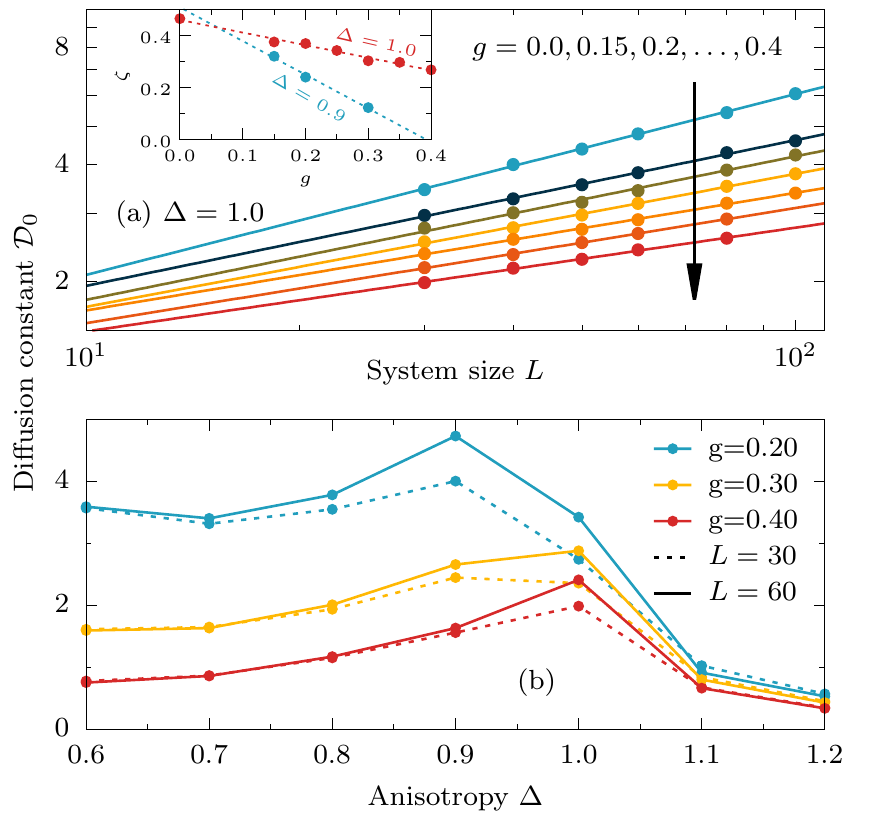}
\caption{ (a) Scaling of NESS diffusion constant vs. $L$ for various perturbations $g =0.15-0.4$, including the unperturbed result $g=0$, fitted with the power laws ${\cal D_0} \propto L^\zeta$ with different $\zeta = [0.24 - 0.5]$. In the inset, fitted $\zeta$ for different $g$ are shown for $\Delta=1.0$ and $\Delta=0.9$.
(b) Diffusion constant ${\cal D}_0$ vs. perturbation $\Delta$ for isotropic perturbation of different strength $g = 0.15 - 0.4$, obtained via NESS method for sizes $L=30, 60$. 
}
\label{fig3}
\end{figure}

For isotropic IBP and $\Delta=1$, Fig.~\ref{fig3}(a) presents the estimated system-size dependence of ${\cal D}_0(L)$ by fitting ${\cal D}_0 \sim L^\zeta$, with exponent $\zeta$ plotted in the inset. Recall that $\zeta$ is related to the dynamical exponent $z$ by $\zeta = 2-z$ \cite{znidaric11,ljubotina17}. In the unperturbed case $g=0$, we get $\zeta=0.46$, complying with the analytically expected KPZ superdiffusion with $\zeta=1/2$. The surprising observation is that also for systems with isotropic IBP, we find a robust signature of superdiffusion, albeit with a smaller exponent $\zeta$, $0.27< \zeta < 0.37$ (but far away from the diffusive $\zeta=0$). Our results agree with previous numerical results \cite{denardis21,roy22} that reported $z=3/2$ scaling for isotropic IBP from the short-time dynamics. Similarly, the recent finite size/time measurements in a cold-atom experiment \cite{wei22} gave an estimate $1.7 < z <1.9$ in weakly perturbed systems. All this suggests that superdiffusion is exceptionally robust against perturbations respecting the SU(2) symmetry, and the expected onset of diffusion would require much larger system sizes (in NESS formalism) and times (for closed systems).

In Fig.~\ref{fig3}(b) we display dependence of ${\cal D}_0(\Delta)$ on $\Delta$ for two different system sizes $L=30,60$ and different isotropic IBP strengths $g=0.2, 0.3, 0.4$. At stronger IBP $g=0.3, 0.4$, we observe a peak in the diffusion constant ${\cal D}_0$ at $\Delta=1$, similarly as in ED calculations at finite magnetization densities $m>0$, Fig~\ref{fig2} and \cite{supmat}. Although at weaker strength $g=0.2$ the peak of ${\cal D}_0(\Delta)$ moves inside the $\Delta<1$ regime, comparing the data for $L=30,60$ makes it apparent that results are only well-converged (with the system size) away from $\Delta \simeq 1$. For this purpose, we repeat the system-size analysis also for $\Delta=0.9$. Apparently, isotropic IBP promotes anomalous ${\cal D}_0 \sim L^{\zeta_{0.9}}$ scaling also for $\Delta = 0.9$, but with $\zeta_{0.9}<\zeta$, see inset of Fig.~\ref{fig3}(a). This explains that with increasing $L$, $\mathcal{D}_0(\Delta=0.9)$ grows slower than $\mathcal{D}_0(\Delta=1.0)$ and the $\Delta\approx 0.9$ peak for $g=0.2$ is only a finite size effect, while we expect that the true thermodynamic peak is at $\Delta=1$. Similarly, we believe that $\Delta<1$ peak at canonical and GC result for the closed system at $m=0$, inset of Fig.\ref{fig2}, is likewise an artifact of small system sizes. To ensure ballistic scaling at $g=0$, $\zeta_{0.9}$ could either cross over to $\zeta_{0.9}\to 1$ at small $g$ or show a discontinous jump at $g=0$.

\noindent{\it Spin stiffness at finite magnetization $m>0$.} 
We finally discuss certain interesting properties of the unperturbed system at finite magnetization that have so far gone undetected. The distinctive feature of the unperturbed ($g=0$) integrable spin chains at finite magnetization is the ballistic spin transport at $T>0$, i.e., finite spin stiffness $D$ in the spin conductivity $\sigma(\omega)=2 \pi D \delta(\omega)+ \sigma_{reg}(\omega)$. In a system with PBC and given $s^z_\mathrm{tot}$, the $D(T\gg1)$ can be evaluated using full exact diagonalization (ED) by calculating all diagonal and degenerate matrix elements $T D = \sum_{\epsilon_n=\epsilon_l} \langle n| j_s | l\rangle^2/(2LN_\mathrm{st})$, with $N_\mathrm{st}$ as the number of many-body states. In the thermodynamic limit, the exact computation of $D$ is possible using the GHD formalism \cite{Ilievski17,bulchandani21}. In the following, we find it convenient to present and discuss the normalized spin stiffness $D^* = D/\chi_0$, with static spin susceptibility $\chi_0 = (1/4-m^2)/T$ providing the conductivity sum rule $\chi_{0}=\tfrac{1}{\pi} \int \sigma(\omega){\rm d}\omega$.

\begin{figure}[tb]
\includegraphics[width=1.0\columnwidth]{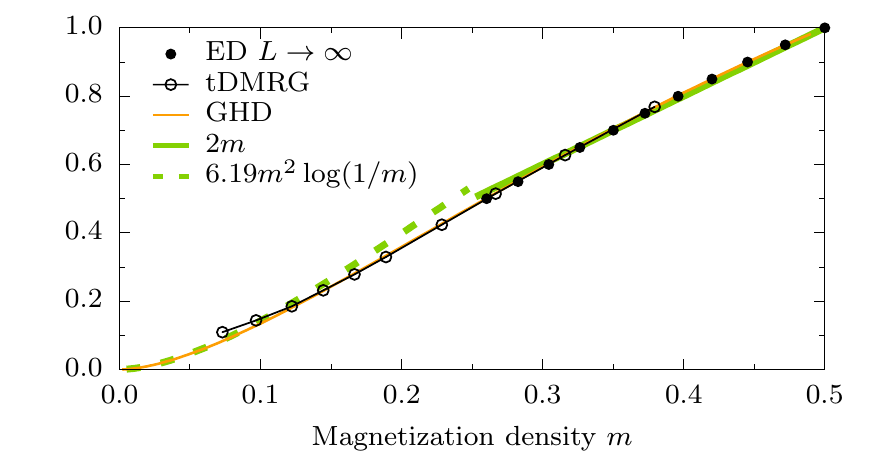}
\caption{(a) Normalized spin stiffness $D^*$ vs. magnetization density $m$ obtained from (i) extrapolated $L\to\infty$ ED data \cite{supmat} (solid black points), (ii) tDMRG result Ref.~\cite{medenjak17} (open black points), and (iii) GHD exact calculation (orange line). Simple linear relation $D^* = 2 |m|$ (green solid line) at large magnetization and $D^*(m)=6.19 m^2 \log(1/|m|)$ (green dashed line) at small magnetization is also presented.}
\label{fig4}
\end{figure}

We concentrate here on the most interesting isotropic point $\Delta=1$, relegating the discussion of other regimes to \cite{supmat}. In Fig.~\ref{fig4}, we show (i) ED results, with extrapolations to $L \to \infty$, (ii) tDMRG result from Ref. \cite{medenjak17}, and (iii) GHD exact results \cite{Ilievski17,bulchandani21}, obtained as detailed in \cite{supmat}. For large magnetizations, the observed approximate slope of $2|m|$ can be, from the viewpoint of GHD, accurately captured by the contributions of magnons and two-magnon bound states \cite{supmat}. Despite larger bound states becoming increasingly important at lower magnetization, it turns out, unexpectedly, that for $m \gtrsim 0.3$, contributions conspire to a nearly linear curve. As elaborated in \cite{supmat}, upon approaching close to half filling ($m=0$), the behavior crosses over to the theoretically established \cite{ilievski18,bulchandani21} anomalous non-analytic scaling $D^*(m) \sim m^2 \log(1/|m|)$, signaling the onset of superdiffusion at $m=0$. Despite different order of limits ($\lim_{L\to \infty} \lim_{t\to \infty} $), extrapolation of our ED results within the accessible range is found in good agreement with the tDMRG data and GHD. At lower magnetization densities, tDMRG values fall on top of the GHD curve until they also depart from it due to finite time limitations.

\noindent{\it Conclusions.}
In this Letter, we show that the anomalous behavior of transport in isotropic Heisenberg chain is not limited to the unperturbed (i.e., integrable) model nor to zero average magnetization $m=0$ (zero external magnetic field), but manifests itself over the entire range of $m$ and at unexpectedly strong perturbations. For isotropic perturbations and $\Delta=1$, we observe anomalously large diffusion constants, with non-monotonic dependence of $\mathcal{D}_0(\Delta)$ on $\Delta$ for all magnetizations, and with the peak at $\Delta=1$ becoming more pronounced at larger $m$. Anisotropic perturbations yield much smaller diffusion constants $\mathcal{D}_0(\Delta)$ with monotonous $\Delta$ dependence. This suggests that within the isotropic Heisenberg model, the quasi-particles are less severely affected by isotropic than anisotropic perturbations. Whether large diffusion constants observed in the isotropically perturbed model can be explained from the broadening of the unperturbed spin stiffness associated with the quasiparticle decay channels or perhaps arise from the dissolution of the anomalous (i.e. singular) diffusion constant of the unperturbed isotropic model remains an open future challenge.

The most challenging to discern is the behavior of the perturbed Heisenberg chain at $m=0$. Here, the anisotropic perturbation suppresses the superdiffusion and leads to finite ${\cal D}_0$ well converged with system size. On the other hand, for the isotropic perturbation, even the largest open-system NESS results conform with a superdiffusive scaling with $L$, ${\cal D}_0 \sim L^\zeta$, with exponent $\zeta \in [0.25, 0.4]$ decreasing with increasing perturbation strength $g$. The intriguing conclusion of our findings is that for finite systems, certain features of the unperturbed KPZ superdiffusion with $\zeta=1/2$ remain fairly robust even for moderately strong isotropic perturbations. Moreover, dynamical exponents in the range $z=2-\zeta \in [1.6,1.75]$ are well consistent with recent experiments on spin superdiffusion in cold-atom lattices, where the perturbation to the Heisenberg chain is added via the exchange between neighboring chains \cite{wei22}. As far as non-monotonous dependence of $\mathcal{D}_0(\Delta)$ is concerned, our open system computations with $m=0$ corroborate the $m>0$ MCLM results obtained in closed systems. 

Although the present study is restricted to the staggered exchange perturbation, we have found a similar dichotomy of anisotropic vs. isotropic perturbations in other examples via MCLM in closed systems, e.g., the next-neighbor exchange, bringing our analysis even closer to actual experiments \cite{wei22}.

\begin{acknowledgments}
The authors thank M. \v Znidari\v c for useful discussions on open-systems results. M.M. acknowledges the support by the National Science Centre, Poland via projects 2020/37/B/ST3/00020. P.P. acknowledges the support by the project N1-0088 of the Slovenian Research Agency. E.I. acknowledges the support by the project N1-0243 of the Slovenian Research Agency. Z.L. and S.N. acknowledge the support by the projects J1-2463 and P1-0044 program of the Slovenian Research Agency. 
\end{acknowledgments}

\bibliography{manuisotrop}
\newpage
\phantom{a}
\newpage
\setcounter{figure}{0}
\setcounter{equation}{0}
\setcounter{page}{0}

\renewcommand{\thetable}{S\arabic{table}}
\renewcommand{\thefigure}{S\arabic{figure}}
\renewcommand{\theequation}{S\arabic{equation}}
\renewcommand{\thepage}{S\arabic{page}}

\renewcommand{\thesection}{S\arabic{section}}

\onecolumngrid

\begin{center}
{\large \bf Supplemental Material:\\
Spin diffusion in perturbed Heisenberg spin chain}\\
\vspace{0.3cm}
S. Nandy$^{1}$, Z. Lenar\v ci\v c$^{1}$, E. Ilievski$^{2}$, M. Mierzejewski$^{3}$, J. Herbrych$^{3}$ and P. Prelov\v{s}ek$^{1}$\\
$^1${\it Department of Theoretical Physics, J. Stefan Institute, SI-1000 Ljubljana, Slovenia} \\
$^2${\it Faculty for Mathematics and Physics, University of Ljubljana, Jadranska ulica 19, 1000 Ljubljana, Slovenia} \\
$^3${\it Department of Theoretical Physics, Faculty of Fundamental Problems of Technology, \\ Wroc\l aw University of Science and Technology, 50-370 Wroc\l aw, Poland}\\
\end{center}

In the Supplemental Material, we present additional numerical results for the spin diffusion constant in the perturbed Heisenberg chain. In particular, we provide more details on the open-systems NESS results, system size and perturbation strength scalings, and additional results from closed grand canonical calculation. The last section highlights the calculation of spin stiffness using generalized hydrodynamics.

\vspace{0.6cm}

\twocolumngrid

\label{pagesupp}

\subsection{Boundary driven open systems: different transport regimes} \label{app0}
The zero magnetization $m=0$ results reported in the main text are predominantly obtained by inducing spin current through the system by coupling it to the baths at the edges. The evolution of the system's density matrix $\rho(t)$ is then governed by the Lindblad master equation,
\begin{eqnarray}
\partial_{t}\rho = -i[H, \rho] + \hat{\mathcal{D}} \rho,
\label{S1}
\end{eqnarray}
where $H$ denotes the system Hamiltonian and $\hat{\mathcal {D}}$ the dissipator, which is expressed in terms of the Lindblad operators $L_{k}$ as $\hat{\mathcal{D}}\rho=\sum_{k}L_{k}\rho L_{k}^{\dagger}-\frac{1}{2}\{L_{k}^{\dagger}L_{k}, \rho \}$. For our problem, $H$ is the XXZ Hamiltonian described in Eq.~\eqref{xxz} in the main text, while Lindblad operators $L_{k}$ act only on the boundary spins and are given by $L_1=\sqrt{1+\mu}S_{1}^{-}, L_2 = \sqrt{1-\mu}S_{1}^{+}, L_3= \sqrt{1-\mu}S_{L}^{-}, L_4 = \sqrt{1+\mu}S_{L}^{+}$. Bias $\mu$ imposes a finite boundary local magnetization and spin current in the non-equilibrium steady state (NESS), $j_{ss}=\text{Tr}[\rho_{ss}j_{s}]$. To evolve the density matrix towards the steady state $\rho_{ss}$, we employ the time-evolving block decimation for vectorized density matrices. In particular, we use the fourth order TEBD with a time step $dt=0.2$, bond dimension $\chi \sim 180$, and bias $\mu \sim 0.001$.

\begin{figure}[!b]
\includegraphics[width=1.0\columnwidth]{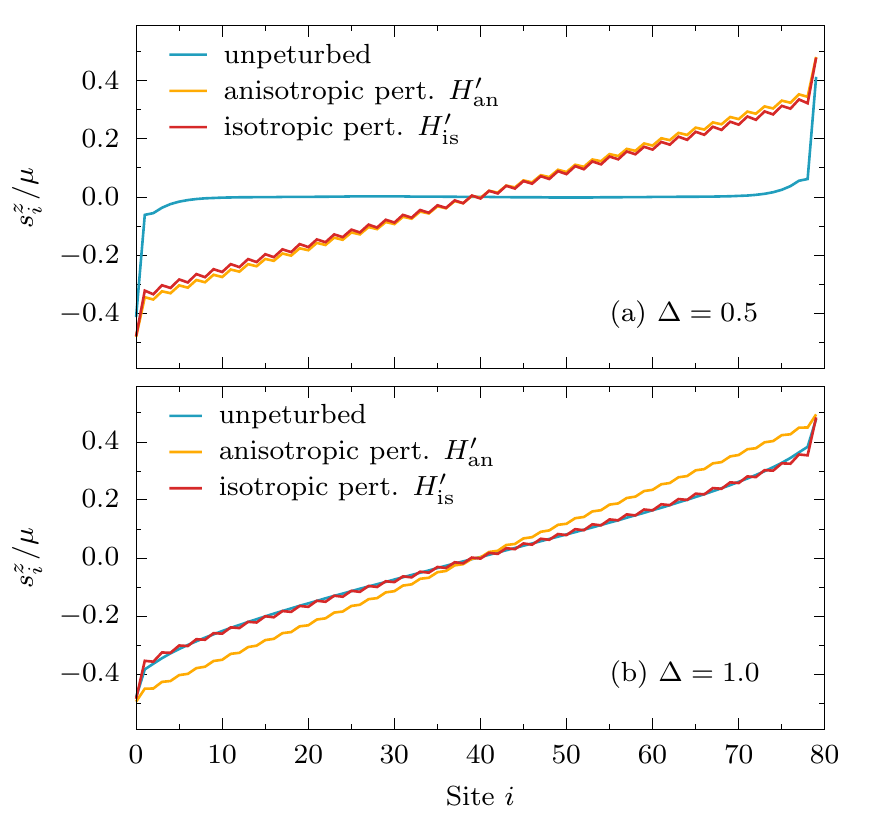}
\caption{Spin profiles $\text{Tr}(S^z_i \rho_{ss})/\mu$ for different $\Delta=0.5, 1.0$ and $g=0, 0.3$ parameters. 
(a) Shows unperturbed, isotropically and anisotropically perturbed ballistic case for $\Delta=0.5$. Here, both IBPs have comparable effect. (b) Shows unperturbed, isotropically and anisotropically perturbed superdiffusive case for $\Delta=1.0$. Unperturbed profile resembles $\arcsin (x)$ profile. For anisotropic perturbation, spin profile is linear with negligible jump at the edges. Isotropic perturbation has much more anomalous effect, leading to pronounced jumps and only approximately linear bulk profile. $L=80$.
}
\label{figS1}
\end{figure}

In Fig.~\ref{figS1}, we show the NESS magnetization profiles for different choices of $\Delta$ and types of perturbations, as described in the main text. The upper panel indicates how the flat profile in the easy-plane regime, characteristic of ballistic transport, gives way to a linear diffusive profile coexisting with the boundary jumps caused by finite system sizes. Here isotropic and anisotropic IBP yield similar effects. Staggering effects in the magnetization profile are related to the staggering in our perturbations. In the lower panel, we perform the same analysis for $\Delta=1$, showing how the superdiffusive profile with a tentative form $s_i^z \sim \frac{\mu}{\pi} \arcsin(-1+2\frac{i-1}{L-1})$, $s^z_i = \text{Tr}(S^z_i \rho_{ss})$ \cite{znidaric11} gets modified by our perturbation. While anisotropic IBP yields a mostly linear diffusive-like profile with small jumps at the edges, the profile for isotropic perturbation has much more pronounced jumps and only an approximately linear bulk profile.

\begin{figure}[t]
\includegraphics[width=1.0\columnwidth]{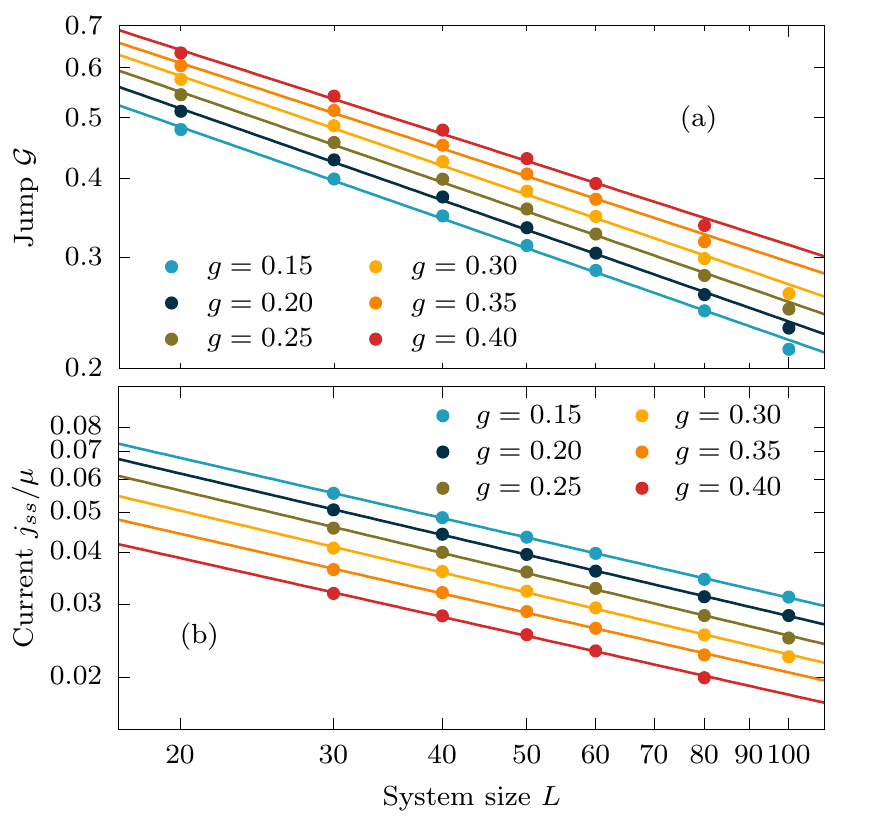}
\caption{Open-system NESS results vs. $L$ for $m=0$ and isotropic perturbations $g=0.15 - 0.4$: (a) the relative strength of boundary jump ${\cal G}$ with corresponding fits to power law ${\cal G} \sim L^{-\zeta}$ with $\zeta \approx 0.5$, (b) the steady state current $j_{ss}$ with power-law fits $j_{ss} \sim L^{-\gamma}$ and $\gamma \approx 0.5$.
}
\label{figS2}
\end{figure}

In Fig.~\ref{figS2}(a) we show the relative value for the magnetization jump
\begin{equation}
{\cal G}= \frac{s^z_2-s^z_1}{\mu}=\frac{s^z_L-s^z_{L-1}}{\mu}\,,
\end{equation}
at the boundaries for the isotropic IBP. As a function of system size, ${\cal G}$ is decaying approximately as ${\cal G} \sim L^{-\zeta'}$ with $\zeta' \approx 0.5$. Such a dependence seems to be another indication of the remnant signatures of superdiffusion in the presence of isotropic IBP and most probably correlates with roughly superdiffusive (KPZ) scaling of the steady state current $j_{ss}\sim L^{-\gamma}$ with $\gamma \approx 0.5$, Fig.~\ref{figS2}(b), for all IBP strengths considered. While in boundary driven setups, transport is often deduced directly from the $j_{ss}(L)$ scaling, as ballistic for $\gamma=0$, diffusive for $\gamma=1$, etc., we note here that to avoid additional misleading finite-size effects, transport properties should be determined from the diffusion constant extracted from the bulk of the system, as we explain in the main text. 

\subsection{Heisenberg chain: anisotropic perturbation } \label{app1}

\begin{figure}[!htb]
\includegraphics[width=1.0\columnwidth]{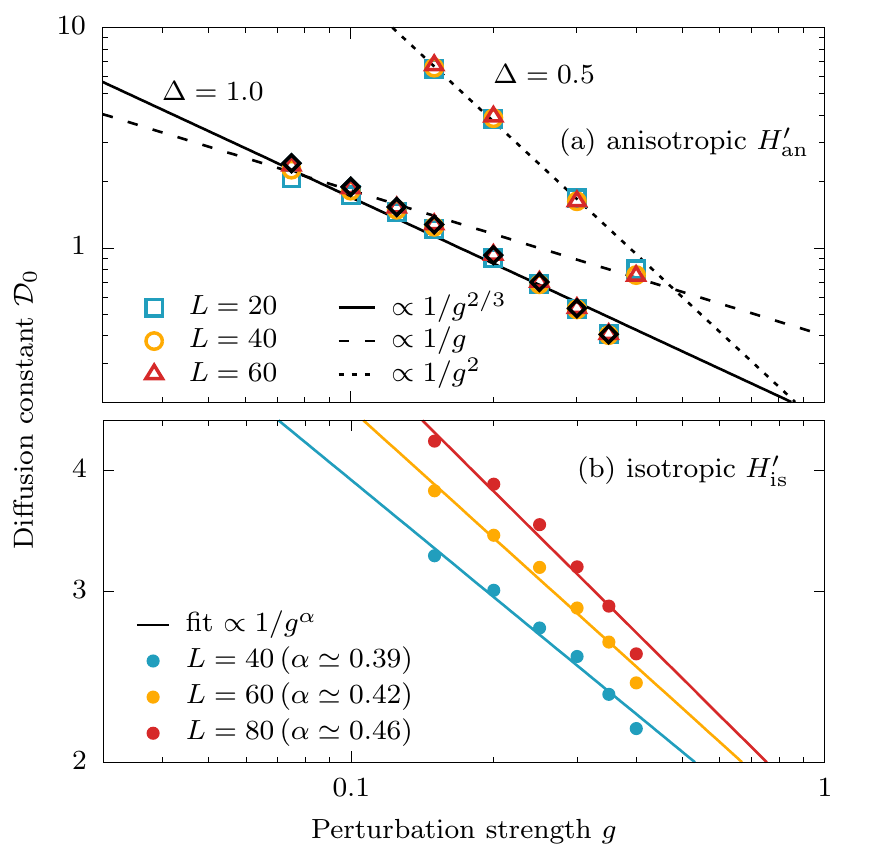}
\caption{(a) Diffusion constant ${\cal D}_0$ vs. anisotropic perturbation strength $g$ at $m=0$ for $\Delta=0.5$ and $\Delta=1$, calculated via NESS for open systems with lengths $L = 20,40,60$. For $\Delta=0.5$, results are fitted with scaling $1/g^\alpha$, $\alpha=2$. For $\Delta=1$, we apparently cannot reach large enough $L$ and small enough $g$ to observe any $1/g^\alpha$ scaling, in particular analytically anticipated $\alpha=2/3$ \cite{znidaric20}. (b) Diffusion ${\cal D}_0$ vs. isotropic perturbation strength $g$ for $\Delta=1$, as obtained via open-system NESS for different $L = 40 - 80$.
}\label{figS3}
\end{figure}

In Fig.~\ref{figS3}(a), we present the scaling of ${\cal D}_0$ with the anisotropic IBP strength $g$, as obtained via open-system NESS method for different sizes $L$ and two $\Delta =0.5, 1.0$. The behavior at $\Delta =0.5$ confirms the standard perturbation-theory scaling ${\cal D}_0 \propto 1/g^2$ scaling. On the other hand, at the isotropic point $\Delta=1$, we do not see the analytically anticipated ${\cal D}_0 \propto 1/g^{2/3}$ scaling \cite{znidaric20} yet, presumably because we cannot access small enough $g$ and correspondingly large $L$.

\subsection{Heisenberg chain: isotropic perturbation } \label{app2}

\begin{figure}[b]
\includegraphics[width=1.0\columnwidth]{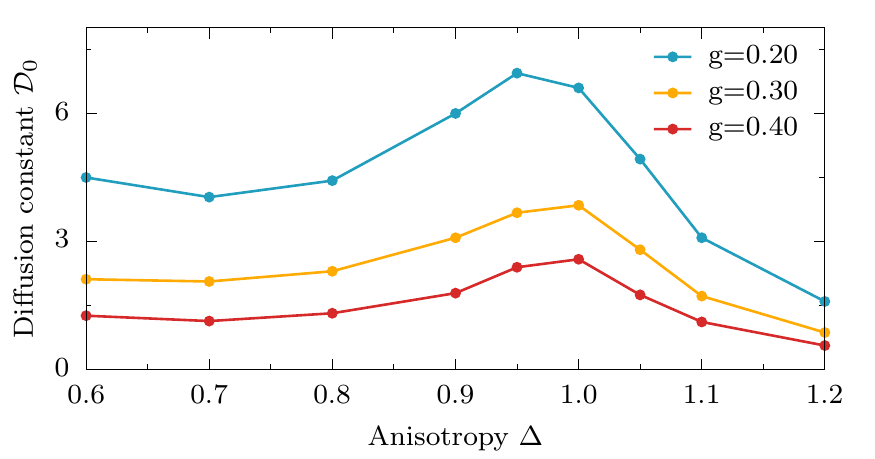}
\caption{ Diffusion constant ${\cal D}_0$ vs. anisotropy $\Delta$ for the different isotropic perturbation strengths $g= 0.2, 0.3, 0.4$, evaluated as the GC average over magnetization sectors $m$, calculated with MCLM on $L=28$ system.}
\label{figS4}
\end{figure}

Isotropic perturbations are considerably more delicate for numerical analysis and interpretation. Our first observation is that for closed-system MCLM at fixed $m=0$, canonical and GC results are much further apart. The GC results in Fig.~\ref{figS4} show dependence of ${\cal D}_0$ with $\Delta$, obtained via MCLM in $L=28$ system for three strengths $g =0.2, 0.3, 0.4$. These results can be compared with the canonical $m=0$ one for the same system in Fig.~\ref{fig2} of the main text for $g=0.2$. While away from $\Delta=1$ both approaches yield similar results, the GC calculation reveals a maximum near $\Delta=1$ for all $g$, in contrast with much smaller canonical $m=0$ values. For the canonical calculation, the shifting of the maxima from $\Delta =1$ for $g \geq 0.3$ to $\Delta < 1$ at $g<0.3$ (for given size $L=28$), as well the overall $\mathcal{D}_0(\Delta)$ variation, is well consistent with the open-system NESS results, presented in Fig.~\ref{fig3}(a) of the main text also for larger $L$.

Finally, we show in Fig.~\ref{figS3}(b) also the analysis of NESS results for the variation of ${\cal D}_0$ at $\Delta=1$ with the isotropic perturbation strength $g$, presented at several lengths $L =40 -80$. We do not see any clear $g$ scaling for system sizes and perturbation strengths considered.

\subsection{Spin stiffness in the unperturbed Heisenberg model at finite magnetization} \label{app3}
Ballistic spin transport at $T>0$ is a distinctive feature of the unperturbed ($g=0$) integrable spin chains at finite magnetization, reflecting in a finite spin stiffness $D$ in the spin conductivity 
\begin{equation}\label{eq::sigma}
\sigma(\omega)=2 \pi D \delta(\omega)+ \sigma_{reg}(\omega)
\end{equation}
Spin stiffness $D(T\gg1)$ is with exact diagonalization (ED) evaluated from 
\begin{equation}
D = \frac{1}{T 2LN_\mathrm{st}}\sum_{\epsilon_n=\epsilon_l} \langle n| j_s | l\rangle^2, 
\label{eqn:DL}
\end{equation}
with $N_\mathrm{st}$ as the number of many-body states. Alternatively, the exact result for $D$ can be, in the thermodynamic limit, calculated with the GHD formalism \cite{castro-alvaredo16,bertini16}, as described in \cite{Ilievski17,bulchandani18,bulchandani21}. In the high-$T$ limit, $D$ as a function of magnetization density $m$ (related to the GC chemical potential $h$ coupling to the global spin projection $s^{z}_{\rm tot}=\sum_{i}S^{z}_{i}$ via $m(h)=\tfrac{1}{2}\tanh{(h/2)}$), reads explicitly \cite{Ilievski17,IlievskiPRB17,bulchandani21}
\begin{equation}
D = \frac{1}{2T}\sum_{s=1}^{s_{\text{max}}} \int d\theta 
\rho_s^{\textrm{tot}}(\theta) 
n_s \, (1- n_s) (v_s^{\text{eff}}(\theta) m_s^{\text{dr}})^2,
\label{eqn:exact_stiffness}
\end{equation}
normalized to match the convention from review \cite{bertini21}, consistent with definition \eqref{eq::sigma}. In this hydrodynamic formula~\footnote{Unlike equation \eqref{eqn:DL}, which is defined for finite systems, $D$ defined in equation \eqref{eqn:exact_stiffness} is only valid in thermodynamic systems (with the $t\to \infty$ limit taken at the end).}, the sum runs over the contributions from single magnons ($s=1$) and magnonic bound states with $s$ quanta of magnetization ($s>2$), while integration is performed over the entire range of rapidities (parameterized by magnon momenta $p_{s}=p_{s}(\theta)$). Moreover, $\rho_s^{\text{tot}}(\theta)$ denote the total densities of available states for the specie $s$, while $n_s$ are the corresponding Fermi occupation functions (which in the high-$T$ limit become independent of $\theta$), $v_s^{\text{eff}}(\theta)$ are the effective velocities of propagation and $m_s^{\text{dr}}$ their dressed value of magnetization. Note that all the state functions depend nontrivially on $m$. For $\Delta \ge 1$, the spectrum of quasiparticles forms an infinite tower of bound states (with bare magnetization $m_{s}=s$) and hence $s_{\text{max}}=\infty$, whereas for $\Delta<1$ the number of quasiparticle species is always finite, but it depends on $\Delta$ (e.g., for $\Delta=0.5$, $s_{\text{max}}=3$). In the latter case, the exact formulae for the thermodynamic state functions can be found e.g. in \cite{ilievski18,popcorn}.

\begin{figure}[!t]
\includegraphics[width=1.0\columnwidth]{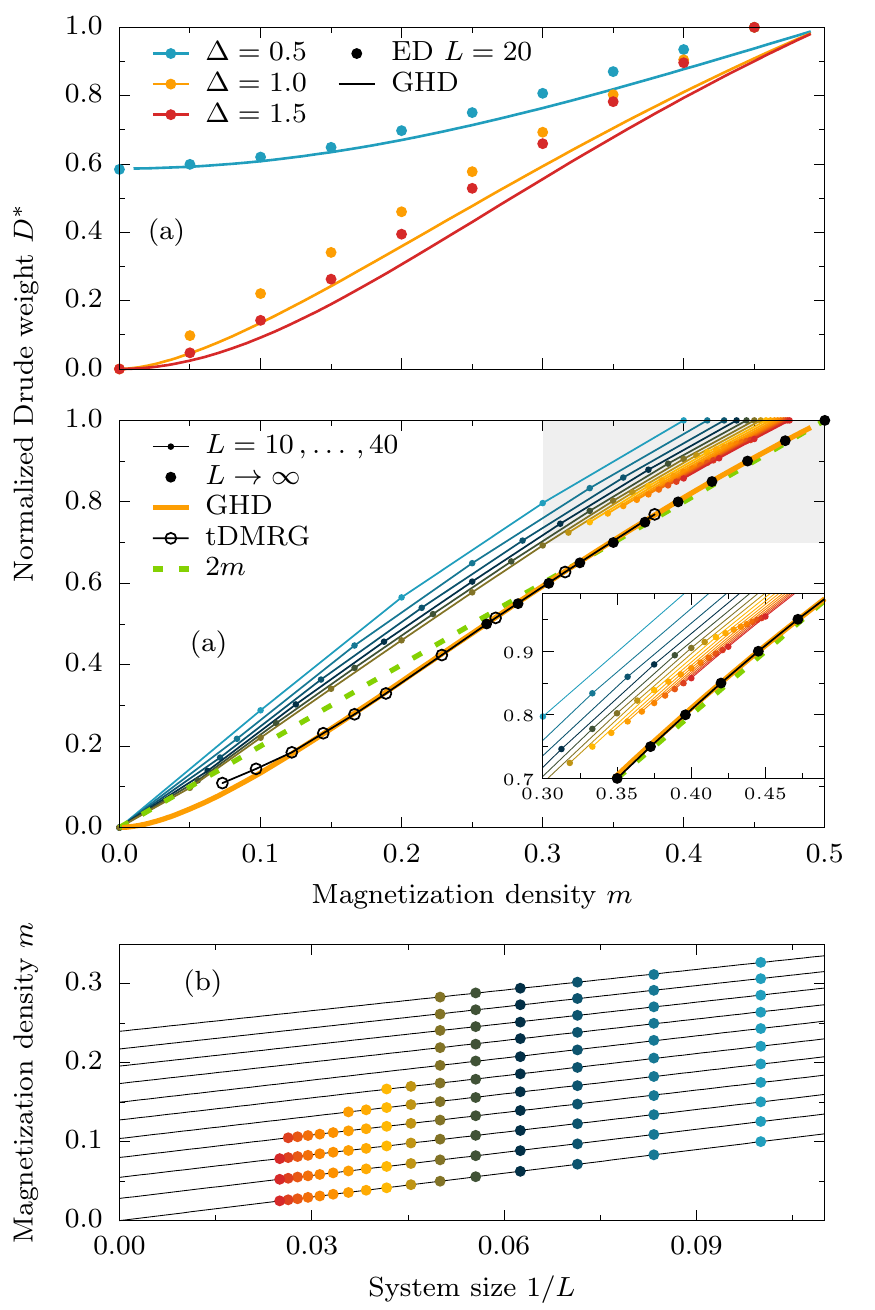}
\caption{(a) Normalized spin stiffness $D^*$ vs. magnetization density $m$ obtained via ED in systems with $L=20$ sites for different anisotropies $\Delta =0.5, 1.0, 1.5$. (b) Finite-size scaling of $D^*$ vs. $m$ for isotropic system $\Delta=1$ for different $L=10,12,\dots,40$, together with extrapolated $L \to \infty$ results (solid black points), tDMRG result Ref.~\cite{medenjak17} (open black points), GHD exact calculation (orange line), and simple linear relation $D^* = 2 |m|$ (green dashed line). Inset depicts detailed data close to $m\sim0.5$ magnetization density. (c) $1/L$ scaling of magnetization density $m$ of interpolated $D^*=0.5,0.55,\dots,1.0$ (top to bottom).}
\label{figS5}
\end{figure}

In the following, we present and discuss the normalized spin stiffness $D^* = D/\chi_0$. Dependence of $D^*(m)$ on $m$ is depicted in Fig.~\ref{figS5}(a), shown for three characteristic regimes corresponding to easy-plane $\Delta<1$, isotropic, and easy-axis regime $\Delta>1$ \cite{herbrych11}. While for $\Delta<1$ we find $D^*(m \to 0) >0$, consistently with ballistic spin transport and partly conserved $j_s$ \cite{prosen11,prosen13}, for $\Delta>1$, $D^*(m \ll 1/2) \propto m^2$ is well captured by the Mazur bound \cite{zotos97}, where the quadratic dependence comes from a large projection of spin current on the conserved energy current. In all cases $D^*(m \to 1/2)=1$ is plausible since the latter reflects dissipationless propagation in the case of low density of excitations (magnons). Quantitative agreement between ED on finite $L=20$ and GHD is evidently rather poor. The reason is different quasi-particle content; for example, at $m=1/2 - 1/L$ magnetization density, ED contains only magnons, while GHD calculation takes into account also bound states with $s>1$. Therefore, a direct comparison to GHD only makes sense by subsequently performing the $L \to \infty$ extrapolation of ED results, as done in Fig.~\ref{figS5}(b,c) for $\Delta=1$. In the accessible range of ED we find perfect agreement, despite the different order of limits $t \to \infty$ and $L\to\infty$ taken in the two approaches. As we elaborate below, dependence of $D^{*}(m)$ on $m$ in the isotropic spin chain consists of two opposite regimes: a roughly linear dependence $D^*(m)\simeq 2 |m|$ for $m\gtrsim 0.3$, and a non-analytic scaling $D(m)\approx a\,m^{2}\log{(1/m)}$ for small $m$; by numerical evaluation of \eqref{eqn:exact_stiffness} we find $a \simeq 6.19$.

We proceed by explaining the numerically observed slope $D^*(m)\simeq 2 |m|$ at large magnetization density. With aid of numerical evaluation, we find that in the $h\to \infty$ limit, the rapidity integrals can be for $s \lesssim 50$ accurately approximated as $\int d\theta \rho_s^{\textrm{tot}}(\theta,m) v_s^{\text{eff}}(\theta, m)^2 \approx s^{-2}/2$, yielding a compact approximation for spin stiffness
$D(m) \approx  \sum_{s=1}^{s_{max}} \tilde{D}_s(m)$ with
\begin{equation}
\tilde{D}_s(m) \equiv \frac{1}{2T} \frac{1}{2s^2}  
n_s(m) \, (1- n_s(m)) m_s^{\text{dr}}(m)^2. 
\end{equation} 
By differentiating this result with respect to $m$, we readily retrieve the anticipated slope of $D^{*}(m)$, namely
\begin{equation}
\partial_m D^{*}(m)\big|_{m=\frac{1}{2}}
= \partial_m \frac{\sum_{s=1}^{s_{max}} \tilde{D}_s(m)}{(1/4-m^2)} \Big|_{m=\frac{1}{2}} = 3-1=2.
\end{equation}
Numerically we find that only magnons ($s=1$) and bound states ($s=2$) are relevant, contributing slopes $3$ and $-1$, respectively, while all other $s>2$ terms have a vanishing slope at $m=1/2$. Nonetheless, $s=1,2$ species feel the presence of the full quasi-particle content implicitly via dressing. It is rather remarkable that the same slope of $2m$ persists approximately even at lower magnetization densities until $m\gtrsim 0.3$, despite higher bound states ($s>2$) contribute appreciably in this regime. The exact infinite sum expression for $D$ is unfortunately difficult to handle analytically, and thus we currently lack a simple effective model to capture this behavior.

In the proximity of half filling, i.e. for small $m$, the spin stiffness develops a non-analytic form $D^{*}(m)\sim m^2 \log(1/|m|)$, already inferred in previous works \cite{ilievski18,bulchandani21}. This type of non-analytic behavior is inherently tied to the onset of superdiffusion. To this end, we briefly rehash the arguments of \cite{ilievski18,gopalakrishnan19,bulchandani21}. By imposing a cutoff $s_{max}=s_{\star}$, the curvature $C$ of spin stiffness $D(h)$ for small $m\sim h$, defined via $D(h) \sim C\,h^{2}/2$, is found to diverge as $C\sim \log{(s_{*})}$ \cite{ilievski18}, readily implying non-analytic behavior at $h=0$, $D(h)\sim h^{2}\log{(s_{*})}$. It then remains to deduce the cutoff dependence $s^*(h)$ on chemical potential $h$. This can be directly inferred from dressed magnetization $m^{\rm dr}_{s}(h)$, which experiences a crossover from $m^{\rm dr}_{s}\sim hs^{2}$ at $sh \ll 1$ to $m^{\rm dr}_{s}(h)=m_{s}=s$ at $sh\gg 1$ \cite{gopalakrishnan19}; cutoff $s^*\sim 1/|h| \sim 1/|m|$ thus effectively separates light species from giant magnons that importantly contribute at small $h$. At small $h\sim m$, spin stiffness thus shows $D(m) \sim m^2 \log{(1/|m|)}$ dependence.

Accordingly, since the spin diffusion constant $\mathcal{D}_{I}$  at $h=0$ diverges linearly with the cutoff $s_{*}$ as $\mathcal{D}_{I}\sim s_{*}$, it diverges algebraically as $\mathcal{D}_{I}(m)\sim 1/|m|$. The dynamical exponent $z=3/2$ (associated with the KPZ superdiffusion) can be then deduced from the growth in the time domain \cite{gopalakrishnan19}: assuming $\mathcal{D}_{I}(t)\sim t^{\alpha}\sim s^*$, using the scaling relation for the dressed velocities $v^{\rm eff}_{s}\sim 1/s$ at large $s$, and equating the anomalous lengthscale $v^{\rm eff}_{s}t\sim t/s^* \sim t^{1-\alpha}$ with $\sqrt{\mathcal{D}_{I}(t)t}\sim t^{(\alpha+1)/2}$, one deduces $\alpha=1/3$. The variance $\sigma_x^{2}(t)$ thus exhibits asymptotic growth $\sigma_x^{2}(t)\sim \mathcal{D}_I(t)t\sim t^{4/3}=t^{2/z}$, implying $z=3/2$.

\end{document}